\documentclass[runningheads]{llncs}
\usepackage[T1]{fontenc}
\usepackage{amssymb}

\usepackage{newtxtext}       %
\usepackage[varvw]{newtxmath}       
\usepackage{amsmath,booktabs,array, tabularx}
\usepackage{xcolor}
\usepackage{tikz}
\usepackage{pgfplots}
\usepackage{lineno}

\usepackage{hyperref}
\hypersetup{
  colorlinks = true,
  citecolor = blue,
  linkcolor = blue,
  urlcolor = blue
}
\usepackage{orcidlink}

\newcommand{\obar}[1]{\overline{#1}}
\newcommand{\lit}{\ell}
\newcommand{\litneg}{\obar{\lit}}
\newcommand{\xpos}{x}
\newcommand{\xneg}{\obar{x}}

\newcommand{\zneg}{\obar{z}}
\newcommand{\mneg}{\obar{m}}
\newcommand{\cadd}{\textrm{a}}
\newcommand{\cdel}{\textrm{d}}
\newcommand{\entails}{\vDash}
\newcommand{\unitprop}{\mathrel{\vDash_{1}}}
\newcommand{\assignment}{\tau}
\newcommand{\eassignment}{\assignment^{*}}
\newcommand{\drat}{\textsc{DRAT}}
\newcommand{\lvar}{\mathit{var}}
\newcommand{\vars}[1]{\lvar\!\left(#1\right)}
\newcommand{\lits}[1]{\mathit{lit}\!\left(#1\right)}
\newcommand{\CC}{\mathcal{C}}
\renewcommand{\AA}{\mathcal{A}}

\newcommand{\zdata}{Z_{\text{prob}}}
\newcommand{\zaux}{Z_{\text{aux}}}

\usepackage{graphicx}
\begin{document}
\title{Apply All, Prove Once:\\ Embedding Parameterized Lemmas in DRAT}
\titlerunning{Parameterized Lemmas in DRAT}
%
\author{
  Randal E. Bryant~\orcidlink{0000-0001-5024-6613}
  \and Zachary Battleman~\orcidlink{0009-0005-3471-329X}
  \and Marijn J. H. Heule~\orcidlink{0000-0002-5587-8801} 
}
%
\authorrunning{Bryant, Battleman, and Heule}
%
\institute{Computer Science Department \\Carnegie Mellon University, Pittsburgh PA 15213 USA}
%
\maketitle              
\begin{abstract}

  The deletion resolution-asymmetric-tautology (DRAT) proof
  format is widely used to express unsatisfiability proofs of
  Boolean formulas.  Our DRATL format extends DRAT with parameterized
  lemmas, allowing a single proof to succinctly capture the common
  case where multiple instances of a pattern occur in a formula,
  differing in the identities and phases of the variables.  Proofs
  expressed in DRATL can be hierarchical, with the proof of one lemma
  containing instantiations of others.

  Our elaborator program expands a DRATL proof containing multiple,
  nested lemmas into a standard DRAT proof.  Most significantly, the
  elaboration yields compact proofs, where a single sequencing of the
  proof steps for each lemma covers all of its instantiations.
  Several case studies demonstrate that lemmas can provide a powerful
  method to generate short proofs for challenging problems.

%
%

\end{abstract}

\section{Introduction}

Proof generation has had a profound impact on the progress and
application of Boolean satisfiability (SAT) solvers over the past 25
years.  Early developers discovered bugs in their solvers well after
deployment, masked by the minimal output generated when the solver
states that a formula is unsatisfiable~\cite{goldberg:date:2003,zhang:date:2003}.  By requiring the
solver to generate a detailed proof of unsatisfiability, such
occurrences were readily detected.  Even if the solver contains
flaws, the successful generation and checking of a proof demonstrates that the
particular formula being evaluated is indeed unsatisfiable.

Proof generation became a requirement for solvers in the main
track of the annual SAT competition in 2016.  Subsequent
competitions have had more consistent judging, and the solvers have
become noticeably more reliable.  It is now rare for a participant to
report an incorrect result, even when evaluating formulas never
previously tested.

Over time, increasingly powerful proof frameworks have been devised,
enabling solvers to handle more challenging problems~\cite{codel:fmcad:2024,heule:nfm:2019,heule:jar:2019}.  A number of
formally verified proof checking tools have been developed, providing
strong guarantees that a particular execution of the solver can be
fully trusted, even if the solver or some of the tools in the chain
from the solver to the checker have
flaws~\cite{codel:fmcad:2024,cruz:cade:2017,heule:itp:2017,tan:sttt:2023}.  The success
of proof generation for SAT solvers has inspired similar efforts in
other domains, including pseudo-Boolean
reasoning~\cite{gocht:cptai:2020}, quantified Boolean formula (QBF) solvers~\cite{niemetz:sat:2012},
satisfiability-modulo-theory (SMT)
solvers~\cite{armand:cpp:2011,barrett:aap:2015}, model counters~\cite{capelli:aaai:2021,fichte:sat:2022},
knowledge compilers~\cite{bryant:jair:2025}, and
hardware model checking~\cite{yu:cav:2021}.

Despite the advances in solvers and proof frameworks, the
generated proofs remain totally flat in their structure.  They
often contain many repetitions of a single pattern of clauses and reasoning steps,
differing only in the identities and phases of the variables.  As a consequence,
the proofs are often very long, containing hundreds of millions of
steps.  The sheer size of the proofs causes challenges in their
generation, storage, and checking.  By contrast, a 
mathematical proof typically has a hierarchical structure, with nested
sets of lemmas capturing core concepts.  Each lemma need only be
proved once, and then it can be applied many times.  The hierarchical
structure yields more compact proofs, leading to shorter checking
times and more manageable storage requirements.

In this paper, we present a framework for adding parameterized lemmas
to the widely used deletion resolution-asymmetric-tautology (DRAT)
proof framework.  Matching the clausal structure of DRAT, each lemma
consists of a set of antecedent clauses, a set of consequent clauses,
and a proof that the consequent can be derived from the antecedent by
a sequence of (augmented) DRAT proof steps.  These three components
are expressed in terms of a set of \emph{parametric} variables that do
not occur in the input formula.  A lemma can be \emph{instantiated}
multiple times, each with a \emph{substitution} defining a mapping
from the parametric variables to problem variables and their
complements.  An instantiation can also include a \emph{qualifier}
clause, with the negation of this clause defining the context under
which the lemma should hold.  Lemmas can be nested, with the proof of
one lemma instantiating others.  Such a structure can dramatically
reduce proof size.  We have developed the DRATL proof format to
describe lemmas, their proofs, and their instantiations by adding a
small set of commands to the standard DRAT file
format~\cite{heule:arxiv:2016}.

Two important features of our work enable our approach to fit within
existing proof-checking environments and to reduce the effort required for proof checking.
First, a proof containing lemmas can be
expanded into a standard DRAT proof.  Our proof \emph{elaborator}
converts proofs in the DRATL format to ones in the DRAT format.  This
enables the proofs to be checked using a well-developed tool chain~\cite{wetzler:sat:2014},
including ones where the final step uses a formally verified checker.
Second, the elaboration maintains the property that each lemma is
proved only once, regardless of the number of instantiations.  The
compactness of the proofs carries over to the efficiency with which
they can be checked.

The key idea behind our elaborator is captured in the phrase ``Apply
all, Prove once.''  The proof for a lemma occurs only after
all of its instantiations have been identified.  The elaborator
introduces extension variables for the parametric variables
encoding the substitutions for all of the instantiations.  It adds a
series of proof steps justifying each clause in the lemma antecedent
in terms of the existing clauses via the different substitutions.
These can include clauses appearing in the proofs of yet-to-be-proved
lemmas in which the current lemma will be instantiated.  The
elaborator then adds the steps of the lemma proof,
enabling the proof to be performed for all instantiations simultaneously.
The elaborator follows the proof with steps adding 
instantiations of the consequent clauses.

In this paper we present our lemma framework and the process by which
the elaborator can expand lemmas to cover all instantiations such that
each proof is carried out only once.  We present arguments for the
correctness of our elaborator, although it is important to note that
the soundness of the generated proof is guaranteed
by the formally verified checker at the end of the verification tool chain.

We provide examples where
lemmas lead to much more compact proofs than the flat proofs
produced by other approaches.
Although some of these problems already have known methods for
generating compact proofs, our lemma framework provides a general and
powerful mechanism for expressing succinct proofs.


In terms of related work, all existing clausal proof frameworks support 
lemmas consisting of single clauses and without parameters.  That is, when a clause is added to the proof, it can
be used in multiple inferences.  The dependency
 structure of the proof forms a directed acyclic graph, rather than a
tree.  This capability is exploited by CDCL SAT solvers, where learned
clauses enable the solver to ignore large regions of the
solution space.  Our ability to have parameters associated with the
lemmas takes this capability to a much higher level, increasing their expressive power.


In most proof systems, parameterized lemmas are expressed in
first-order logic using quantified variables to serve as the
lemma parameters.  The lemma can be proved once and then reused via multiple quantifier instantiations.
Our approach follows a
similar strategy, but without quantification.  By first
encoding all instantiations with a single set of extension variables, one
sequencing of the proof steps suffices.

\section{Background: The DRAT Proof System}

Here we present our notation for Boolean formulas and clausal proofs based on the DRAT proof system.

\paragraph{\bf Boolean Formulas.}
Boolean variable $x$ has literals $\xpos$ and $\xneg$, the complement
of $x$.  For literal $\lit$, its associated variable is denoted
$\vars{\lit}$.  For literal $\lit$ with variable $x = \vars{\lit}$, we
let $\litneg = \xneg$ when $\lit = \xpos$ and $\litneg = \xpos$ when
$\lit = \xneg$.  A clause is a set of literals, representing their
disjunction.  A clause cannot contain both literals $\xpos$ and
$\xneg$ for any variable $x$.  A Boolean formula in conjunctive normal form (CNF) is a set of clauses,
representing their conjunction.  We extend $\lvar$ to clauses and
formulas by saying that $\vars{C}$ is the set of all variables
associated with the literals in $C$, and $\vars{F}$ is the union of
the sets of all variables associated with the clauses in $F$. Moreover, given a set of 
variables $X$, define $\lits{X} = X \cup \{\obar{x} \mid x \in X\}$. We write $\lits{F}$ to mean
$\lits{\vars{F}}$. Lastly, given two sets of literals, $\lits{X}$ and $\lits{Y}$, a substitution $\sigma : \lits{X} \to \lits{Y}$
is a mapping with the additional condition that $\sigma(x) = \lnot \sigma(\obar{x})$.
Following
common convention, we use combinations of set notation and logical connectives
in writing clauses and formulas.  For example, the formula $F \cup \{ \{ x_1\}, \{\xneg_2, x_3\} \}$
can also be written as $F \land x_1 \land (\xneg_2 \lor x_3)$.

For variables $X = \vars{F}$, a \emph{truth assignment} $\assignment$ maps variables in $X$ to $\{0,1\}$.  It is extended
to literals by defining $\assignment(\xneg) = 0$ when $\assignment(x) = 1$, and $\assignment(\xneg) = 1$ when $\assignment(x) = 0$.
Clause $C$ is \emph{satisfied} by $\assignment$ when some literal
$\lit \in C$ has $\assignment(\lit) = 1$.  Formula $F$ is satisfied by $\assignment$ when
every clause $C \in F$ is satisfied by $\assignment$.
Formula $F$ is
\emph{satisfiable} when it is satisfied by some assignment $\assignment$, and
it is \emph{unsatisfiable} otherwise.  The empty clause having no
literals is written $\bot$.
It cannot be satisfied.  We also use $\bot$ to denote any formula containing the empty clause.
In a similar fashion, we use $\top$ to denote a formula with no clauses; it is trivially satisfied.
Two formulas $F$
and $G$ are \emph{equisatisfiable} when either both are satisfiable or
both are unsatisfiable.
Formula $F$ \emph{entails} formula $G$, written $F \entails G$, when every satisfying assignment of $F$ also satisfies $G$.

\paragraph{\bf Clausal Proofs.}
Clause $C$ is said to be \emph{redundant} with respect to $F$ when the
formulas $F$ and $F \cup \{ C \}$ are equisatisfiable.  A \emph{proof
rule} is a pair of the form $\langle \cadd, C\rangle$, or $\langle
\cdel, C\rangle$, where $C$ is a clause, and ``$\cadd$'' denotes clause addition, while
``$\cdel$'' denotes clause deletion.  Addition rule $\langle \cadd, C\rangle$
is \emph{applicable} for formula $F$ when $C$ is redundant with
respect to $F$.  Its application yields the formula $F' = F \cup \{ C
\}$.  Deletion rule $\langle \cdel, C \rangle$ is always applicable.
It yields the formula $F' = F \setminus \{C\}$.  Observe that addition
rules maintain equisatisfiability, but applying a deletion rule can
turn an unsatisfiable formula into a satisfiable one.

For formula $F$, a \emph{clausal proof} consists of a sequence $\Pi$ of
proof steps $R_1, R_2, \ldots, R_s$, defining a sequence of formulas
$F_0, F_1, \ldots, F_s$, where $F_0 = F$ and each formula $F_i$ is the
result of applying rule $R_i$ to $F_{i-1}$ for $1 \leq i \leq s$.
Each addition step must be applicable.  That is, for rule
$R_i = \langle \cadd, C_i \rangle$,
clause $C_i$ must be redundant with respect to $F_{i-1}$.
For a clausal proof $\Pi$, $\vars{\Pi}$ and $\lits{\Pi}$ are defined analogously to their
formula counterparts.

Proof rule $R_i$ in the sequence is said to be \emph{implication
preserving} when it preserves entailment, i.e., $F_{i-1} \entails F_i$.
This can be guaranteed for a deletion rule by prohibiting the deletion of
clauses from
the original formula $F$.
For
a proof step adding clause $C_i$, implication preservation can be guaranteed
when $F_{i-1} \entails C_i$.

A \emph{clausal proof of unsatisfiability} for formula $F$ is a
proof ending with the proof rule
$R_s = \langle \cadd, \bot \rangle$.  Any formula $F$ having a clausal proof of
unsatisfiability must be unsatisfiable.

For formula $G$, a \emph{clausal derivation} of $G$ from formula $F$
is a sequence $\Pi$ of implication-preserving rules $R_1, R_2, \ldots, R_s$, such that
applying these rules yields a formula $F_s$ with $G \subseteq F_s$.
Such a derivation guarantees the entailment $F \entails G$.

\begin{figure}
\begin{minipage}[t]{0.33\textwidth}
\centering{
\setlength{\tabcolsep}{10pt}
Boolean Formula $F$ $\phantom{\overline z_1}$ \\[0.5ex]
\begin{tabular}{rl}
\toprule
ID & Clause \\
\midrule
1 & $z_1\lor z_2\lor\zneg_3$ \\
2 & $\zneg_1\lor\zneg_2\lor z_3$ \\
3 & $z_2\lor z_3\lor\zneg_4$ \\
4 & $\zneg_2\lor\zneg_3\lor z_4$ \\
5 & $z_1\lor z_3\lor z_4$ \\
6 & $\zneg_1\lor\zneg_3\lor\zneg_4$ \\
7 & $\zneg_1\lor z_2\lor z_4$ \\
\bottomrule
\end{tabular}
} 
\end{minipage}
\begin{minipage}[t]{0.65\textwidth}
\centering{
\setlength{\tabcolsep}{10pt}
Derivation of $G = \zneg_1 \land z_2 \land z_4$ \\[0.5ex]
\begin{tabular}{rcll}
\toprule
ID & Action & Clause & Dependencies \\
\midrule
8 & a & $z_2 \lor z_4$ & 7, 5, 1 \\
9 & a & $z_2$ & 8, 3, 6, 1 \\
  & d & $z_2 \lor z_4$ & 8 \\
10 & a & $z_4$ & 9, 4, 5, 2 \\
11 & a & $\zneg_1$ & 9, 10, 6, 2 \\
\bottomrule
\end{tabular}
} 
\end{minipage}
\caption{Boolean Formula and Clausal Derivation Examples}
\label{fig:formula:example}
\end{figure}

The formula shown on the left side of
Figure~\ref{fig:formula:example}
will serve as an example in our presentation.
A clausal derivation based on that formula is shown on the right.
Each clause is numbered, starting with those in the input and continuing with those added by the proof.
Each step of the derivation either adds or deletes a clause.
The proof adds unit clauses with literals $z_2$, $z_4$, and $\zneg_1$, which we can write as the conjunction $\zneg_1 \land z_2 \land z_4$.
The proof also deletes the intermediate clause $z_2 \lor z_4$.
We describe the column labeled ``Dependencies'' after we have presented the applicable proof rules.

\paragraph{\bf Proof Rules.}
A \emph{unit clause} consists of a single literal $\lit$.  It is
only satisfied by an assignment $\assignment$ having $\assignment(\lit) = 1$.  The
process of \emph{unit propagation} iteratively simplifies a formula,
where each iteration 1) identifies a unit clause consisting of literal
$\lit$, 2) removes all clauses containing $\lit$, and 3) eliminates
literal $\litneg$ from every clause containing it.  This process
iterates until a (unique) fixed point is reached.  We write
$F \unitprop G$ when applying unit propagation to formula $F$ yields
formula $G$.  We can see that unit propagation preserves implication.
That is, any assignment $\assignment$ satisfying $F$ will also satisfy
$G$ when $F \unitprop G$.  Conversely, any formula $F$ for which $F \unitprop \bot$ must be unsatisfiable.

Clause $C = \lit_1 \lor \lit_2 \lor \cdots \lor \lit_k$ satisfies the
\emph{Reverse Unit Propagation} (RUP) proof rule~\cite{vangelder:aim:2008} with respect to formula $F$, when the following process succeeds:
\begin{enumerate}
\item Formula $F$ is augmented by unit clauses consisting of the negated literals of $C$: 
$F' = F \land \litneg_1 \land \litneg_2 \land \cdots \land \litneg_k$
\item Unit propagation yields the empty clause: $F' \unitprop \bot$.
\end{enumerate}
Applying proof rule $\langle \cadd, C \rangle$ to $F$ when $C$ satisfies the RUP proof rule with respect to $F$
preserves implication: any assignment $\assignment$ satisfying $F$ will also
satisfy $C$.

Returning to the example of Figure~\ref{fig:formula:example}, we can now
explain the field labeled ``Dependencies.''  For each clause addition, it
shows the IDs of the clauses that reduce to unit clauses
when performing unit
propagation to validate a RUP clause addition, with the final clause yielding $\bot$.  For the deletion step,
the dependency gives the ID of the deleted clause.  We can see that the
proof is implication preserving, since all additions used the RUP
proof rule, and the only deletion was for a proof clause.

Clause $C$ satisfies the
\emph{Resolution Asymmetric Tautology} (RAT) proof rule~\cite{jarvisalo:ijcar:2012}
with respect to formula $F$ for \emph{pivot} literal $\lit$, where $\lit \in C$,
when for every clause $D \in F$ such that $\litneg \in D$, the following process succeeds:
\begin{enumerate}
\item Form the set of literals $R = C \cup (D \setminus \litneg)$.
\item Either:
  \begin{enumerate}
  \item $R$ contains literals $x$ and $\obar{x}$ for some variable $x$, or
  \item $R$ satisfies the RUP rule with respect to $F$.
  \end{enumerate}
\end{enumerate}

Clause $C$ satisfies the RAT rule with respect to $F$ when it does so
for some pivot $\lit \in C$.  Applying proof rule
$\langle \cadd, C\rangle$ when $C$ satisfies the RAT proof rule with respect to $F$ maintains equisatisfiability, but it may
eliminate assignments that satisfy $F$.  That is, there may be an
assignment $\assignment$ that satisfies $F$, but it fails to satisfy $C$.

The RUP proof rule satisfies a \emph{monotonicity} property such that
if clause $C$ satisfies the RUP rule with respect to formula $F$, then
it also does so for any formula $F'$ such that $F' \supseteq F$.  Such
is not the case for the RAT rule, since it can depend on the absence
of a clause.  As a consequence, the deletion rule is an important
capability for DRAT proofs.  By deleting some clauses that are no
longer needed, it may be possible to perform RAT additions that would
otherwise be prohibited.

A clause $C$ satisfying the RUP rule with respect to $F$ will also
satisfy the RAT rule, with any literal $\lit \in C$ serving as the
pivot.  Given its power, it can be helpful to highlight two special
cases where clause $C$ satisfies the RAT rule with respect to formula
$F$ for pivot $\lit$:
\begin{description}
\item[RUP:] Clause $C$ satisfies the RUP rule with respect to $F$
\item[Blocked Clause:] Every clause $D$ such that $\litneg \in D$ also
  contains a literal $m \not = \litneg$, such that $\mneg \in C$.
\end{description}


The RAT rule enables a proof to add variables not found in the
original formula $F$.  These are known as \emph{extension} variables,
and they provide the DRAT proof system with the capabilities of the powerful extended-resolution proof system~\cite{tseitin:1983}.  Suppose formula $F$ has $X = \vars{F}$ and step $R_i$ in
clausal proof $\Pi$ yields formula $F_i$ having $\vars{F_i} = X \cup Z$.
That is, $Z$ is the set of extension variables in the proof clauses that have been added and not yet deleted.
We say that assignment $\assignment:X \rightarrow \{0,1\}$
satisfies $F_i$ if there is an assignment
$\eassignment:X\cup Z \rightarrow \{0,1\}$ that satisfies $F_i$, such that $\eassignment(x) = \assignment(x)$ for
all $x \in X$.  With this definition, it can be
shown that any application of the RAT proof rule having a literal of an extension
variable as its pivot preserves implication.

To summarize, a clausal derivation can contain any of the following proof steps:
\begin{enumerate}
\item A deletion step that does not delete a clause in the original formula $F$.
\item An addition step based on the RUP proof rule.
\item An addition step based on the RAT proof rule, having the literal of an extension variable as the pivot.
\end{enumerate}

\section{Embedding Lemmas in DRAT}
\label{sec:embedding}
A lemma consists of a set of antecedent clauses $\AA$, a set of
consequent clauses $\CC$, and a clausal derivation $\Pi$ of $\CC$ from
$\AA$. These components are defined over a set of variables $Z$
disjoint from those in the top-level formula $F$.  These variables are
partitioned into variables $\zdata$ that map, via substitutions, back
to literals in $F$ and variables $\zaux$ internal to the lemma proof.
The auxiliary variables enable the lemma proof to employ extended resolution.

The instantiation of a lemma is based on a substitution $\sigma :
\lits{\zdata} \to \lits{F}$ and a set of \emph{qualifying} literals $Q
\subseteq \lits{F}$, such that $\neg Q$ limits the context under which the lemma must hold.
Our procedure derives
the consequent, mapped back to literals in $F$, for multiple instantiations and multiple sets of qualifying literals without needing to apply $\Pi$ 
multiple times.

\begin{figure}[t]
\centering
\setlength{\extrarowheight}{1.5pt}
\begin{tabular}{@{}>{\centering\arraybackslash}p{72.5pt}@{}}
\toprule
Block $a$ \\
\midrule
$a_1\lor a_2\lor\overline{a}_3$ \\
$\overline{a}_1\lor\overline{a}_2\lor a_3$ \\
$a_2\lor a_3\lor\overline{a}_4$ \\
$\overline{a}_2\lor\overline{a}_3\lor a_4$ \\
$a_1\lor a_3\lor a_4$ \\
$\overline{a}_1\lor\overline{a}_3\lor\overline{a}_4$ \\
$\overline{a}_1\lor a_2\lor a_4$ \\
\bottomrule
\end{tabular}
\quad
\begin{tabular}{@{}>{\centering\arraybackslash}p{72.5pt}@{}}
\toprule
Block $b$ \\
\midrule
    $b_1\lor b_2\lor\overline{b}_3 \lor \obar{a}_2$ \\
    $\overline{b}_1\lor\overline{b}_2\lor b_3 \lor \obar{a}_2$ \\
    $b_2\lor b_3\lor\overline{b}_4 \lor \obar{a}_2$ \\
    $\overline{b}_2\lor\overline{b}_3\lor b_4 \lor \obar{a}_2$ \\
    $b_1\lor b_3\lor b_4 \lor \obar{a}_2$ \\
    $\overline{b}_1\lor\overline{b}_3\lor\overline{b}_4 \lor \obar{a}_2$ \\
    $\overline{b}_1\lor b_2\lor b_4 \lor \obar{a}_2$ \\
\bottomrule
\end{tabular}
\quad
\begin{tabular}{@{}>{\centering\arraybackslash}p{72.5pt}@{}}
\toprule
Block $c$ \\
\midrule
$c_1\lor c_2\lor\overline{c}_3\lor b_1$ \\
$\overline{c}_1\lor\overline{c}_2\lor c_3\lor b_1$ \\
$c_2\lor c_3\lor\overline{c}_4\lor b_1$ \\
$\overline{c}_2\lor\overline{c}_3\lor c_4\lor b_1$ \\
$c_1\lor c_3\lor c_4\lor b_1$ \\
$\overline{c}_1\lor\overline{c}_3\lor\overline{c}_4\lor b_1$ \\
$\overline{c}_1\lor c_2\lor c_4\lor b_1$ \\
\bottomrule
\end{tabular}
\quad
\begin{tabular}{@{}>{\centering\arraybackslash}p{72.5pt}@{}}
\toprule
Conflicting clause \\
\midrule
$a_1\lor\overline{b}_2\lor\overline{c}_4$ \\
\ \\ \ \\ \ \\ \ \\ \ \\ \ \\
\bottomrule
\end{tabular}
    \caption{An unsatisfiable formula. Each block, conditioned on the units from the previous, derives the three unit literals corresponding to $\obar{z}_1$, $z_2$, and $z_4$ which
    falsify the final clause.}
    \label{fig:3sat}
\end{figure}
\subsection{The Embedding Procedure}

Based on the notation we have introduced, 
we assume that the sets of variables $Z$, $\zdata$, and $\zaux$ obey the following:
$$Z = \vars{\AA} \cup \vars{\CC} \cup \vars{\Pi} \qquad Z \cap \vars{F} = \varnothing$$ 
$$Z = \zdata \cup \zaux \qquad \vars{\AA}, \vars{\CC} \subseteq \zdata$$
Since $\Pi$ is a clausal derivation, we require that any of its addition steps based on the RAT proof rule have pivot literal $\lit$ such that  $\lit \in \lits{\zaux}$.
Let the consequent $\CC$ consist of the clauses $\CC = \{C_1, C_2, \ldots, C_m\}$.
In the following, we add a number of ``fresh'' variables, each being a previously unused extension variable.

Given $K \geq 2$ applications of the lemma,
let $\sigma_k : \lits{\zdata} \to \lits{F}$ be the desired substitution
and $Q_k$ be the set of qualifying literals for the $k$th instantiation.
All of the clauses in $\sigma_k(\AA)$ must be RUP with respect to $F \land \lnot Q_k$.
For the construction, introduce fresh variable $t$ and fresh variables $s_1, \ldots, s_{(K-1)m}$.  For convenience define $s_0 = \top$ and $s_{Km} = \bot$.
The clause $C$ \emph{guarded} by $t$ is defined to be the clause $\obar{t} \lor C$.
It
behaves exactly like $C$ when $t$ is set to $1$,  and it is satisfied otherwise.
Finally, let $D_{k,j} = Q_k \lor \sigma_k(C_j)$ be the translation of consequent clause $j$ for instantiation $k$.
The procedure is presented in
Figure~\ref{fig:compact} using the notation $[n]$ to represent the set $\{1, \ldots, n\}$. 
\begin{figure}[tb]
\begin{tabularx}{\textwidth}{
  @{}p{0.17\textwidth}
  >{\raggedright\arraybackslash}X
  p{0.21\textwidth}@{}
}
\textbf{Step} & \textbf{Clauses or operation} & \textbf{Rule}
\\
&
&\\
1. Prefixes
&
    For $r = 1, \ldots, (K-1)m$
&
\\
    &
    and $k, j$ such that $r = (k-1)m + j$:
&
\\
&
    $(\overline s_r\lor s_{r-1}),\quad(\overline s_r\lor D_{k, j})$
&
Blocked on $\overline s_r$
\\
&
$(s_r\lor\overline s_{r-1}\lor\overline\ell)$,
for each $\ell\in D_{k,j}$
&
Blocked on $s_r$
\\[8pt]

2. Bindings
&
For $k\in[K]$, increasing, and $z\in \zdata$:
&
\\
&
$\overline z\lor\overline s_{(k-1)m}\lor s_{km}\lor\sigma_k(z)$
&
RAT on $\overline z$
\\
&
$z\lor\overline s_{(k-1)m}\lor s_{km}\lor {\sigma}_k(\obar{z})$
&
RAT on $z$
\\[8pt]
3. Antecedent
&
With fresh $t$:
&
\\
&
$\obar{t} \lor A$ for all $A \in \AA$
&
Blocked on $\obar{t}$
\\
& $t \lor \obar{s}_{(k-1)m} \lor s_{km} \lor Q_k$ for all $k \in [K]$
& RAT on $t$
\\[8pt]

4. Prove
&
For each step $\langle R, P\rangle \in \Pi$:
&
\\
& $\obar{t} \lor P$ if $R = \cadd$
& RAT/RUP as in $\Pi$
\\
&
Delete $\obar{t} \lor P$ if $R = \cdel$
&
Deletion
\\[8pt]

5. Prefix Units
&
For $r = 1,\ldots, (K-1)m$:
&
\\
&
$s_{r}$ 
& RUP
\\[8pt]
6. Consequent
&
For every $k \in [K]$ and $j \in [m]$:
&
\\
&
$D_{k,j}$&
RUP
\\[8pt]
7. Cleanup
&
Delete the added clauses other than the consequent $D_{k,j}$
preserving every clause of $F$.
&
Deletion
\end{tabularx}
\caption{The seven-step procedure to embed a lemma into DRAT. Tautologies are omitted.}
\label{fig:compact}
\end{figure}
We now prove the correctness of the embedding.
\begin{theorem}[Embedding Correctness]
    Suppose that $F, \AA, \CC, \Pi, \sigma_k$, and $Q_k$ satisfy the conditions above
    for all $k$ in $[K]$
    and that all variables introduced are fresh. Then the procedure
    in Figure~\ref{fig:compact} is a valid DRAT derivation
    of $F \cup \{D_{k, j} \mid k \in [K], j \in [m]\}$. 
    \label{thm:embedding}
\end{theorem}
\begin{proof}
    To prove the theorem, we show that the additions and deletions of every step are valid and result in the desired formula at the end.
    \begin{description}
        \item[Prefixes] The first step adds definitional clauses encoding 
            $s_r \leftrightarrow (s_{r-1} \land D_{k, j})$.
            Added in order of increasing $r$, these additions are valid via blocked clause addition.
        \item[Bindings] This step adds two clauses for each problem variable and application.
            Fix an instantiation $k$ and a variable $z \in \zdata$. The clause $\obar{z} \lor \obar{s}_{(k-1)m} \lor s_{km} \lor \sigma_k(z)$
            is RAT on $\obar{z}$. If $k = 1$, then it is added by blocked clause addition. If $k > 1$,
            every clause in the formula which contains the complement, $z$, contains $s_{k'm}$ for some $k' < k$. 
            Thus, the resulting set of literals in the RAT check will have both $s_{k'm}$ and $\obar{s}_{(k-1)m}$. When negated, these
            units propagate down the chain and result in a conflict, satisfying the RAT condition. For the opposing clause,
            $z \lor \obar{s}_{(k-1)m} \lor s_{km} \lor \sigma_k(\obar{z})$, the argument is the same except that the previously
            added clause contains $\obar{z}$ and therefore gives RAT one additional check. The check succeeds because the resulting set of literals 
            contains both $\sigma_k(z)$ and $\sigma_k(\obar{z})$.
        \item[Antecedent] The clauses $\obar{t} \lor A$ for all $A \in \AA$ are added via blocked clause addition. 
            The clauses of the second set are all RAT on $t$. For every $k \in [K]$, the only clauses that are relevant
            to the RAT check for $t \lor \obar{s}_{(k-1)m} \lor s_{km} \lor Q_k$ are the guarded antecedent clauses.
            For each antecedent  clause $A$, the RAT condition requires that $t \lor \obar{s}_{(k-1)m} \lor s_{km} \lor Q_k \lor A$ is RUP.
            Negating this clause results in the two units $s_{(k-1)m}$ and $\obar{s}_{km}$ establishing the bindings
            $z \leftrightarrow \sigma_k(z)$ for $z \in \zdata$. The bindings, along with the units $\lnot A$, establish the 
            units $\lnot \sigma_k(A)$. Along with the units $\lnot Q_k$, this results in a conflict as $\sigma_k(A) \lor Q_k$ is RUP from $F$ by assumption.
        \item[Prove] For each step of $\Pi$, if it is an addition $\langle \cadd, P\rangle \in \Pi$, then $\obar{t} \lor P$ is redundant via the same mechanism as in $\Pi$. There are two cases:
            $P$ is either RAT on an auxiliary variable or RUP. If it is RAT on an auxiliary variable, since that variable only occurs in $\Pi$,
            it is still RAT on that auxiliary variable. If it is RUP, on the other hand, then the addition of $\overline t \lor P$
            is RUP. The unit $t$ activates the antecedent clauses $\AA$ and the earlier guarded steps of $\Pi$, so $P$ is RUP exactly as it is in $\Pi$. 
            If it is a deletion $\langle \cdel, P \rangle$, then the clause $\obar{t} \lor P$ is deleted.
        \item[Prefix Units] 
            Inductively, assume that all units $s_1$ through $s_{i-1}$ have been derived. Choose $k$ such that $(k-1)m < i \leq km$.
            Negating the unit $s_i$ propagates up the prefix chain established in step $1$, yielding the unit $\obar{s}_{km}$.
            Since $(k-1)m < i$, $s_{(k-1)m}$ is already derived, and therefore the bindings from step $2$ 
            activate, yielding $z \leftrightarrow \sigma_k(z)$ for $z \in \zdata$.
            Moreover, since $s_{i-1}$ is already derived, by the prefix clauses in step $1$, negating $s_i$ also derives $\lnot D_{k,j}$ for some $j \leq m$ with $i = (k-1)m + j$.
            Since $Q_k \subseteq D_{k, j}$, the negation of $D_{k, j}$ provides the units in $\lnot Q_k$, which, along with
            $\obar{s}_{km}$ and $s_{(k-1)m}$ propagate $t$. Finally, recognizing that $\sigma_k(C_j) \subseteq D_{k, j}$ and 
            that the bindings $z \leftrightarrow \sigma_k(z)$ are active, the units $t$ and $\lnot C_j$
            conflict with the $j$th consequent of step $4$, $\obar{t} \lor C_j$.
        \item[Consequent] After establishing all of the prefix units, the consequent clauses $D_{k,j}$ for $k < K$ and all $j$ follow immediately from the prefix
            clauses in the first step. Similarly, all of the $D_{K, j}$ are valid since negating each clause yields both $\lnot \sigma_K(C_j)$ and  $\lnot Q_K$.  
            The latter, along with the prefix units, propagates $t$, which activates the guarded consequent clauses of step $4$. In particular,
            $\lnot \sigma_K(C_j)$ and the active bindings $\sigma_K$ conflict with the now active consequent $C_j$.
        \item[Cleanup] Clause deletions are unconditional in DRAT, so we can freely delete all of the scaffolding, yielding
            a DRAT derivation of $F \cup \{D_{k, j} \mid k \in [K], j \in [m]\}$ as desired.
    \end{description}
\end{proof}

\subsection{Embedding Example}
As an example of the embedding, 
consider the formula consisting of 
the conjunction of the four blocks in Figure~\ref{fig:3sat}. 
These are three copies of the formula in Figure~\ref{fig:formula:example},
replacing $z_i$ with $a_i$, $b_i$, and $c_i$, and with 
blocks $b$ and $c$ under the qualifiers $\obar{a}_2$ and $b_1$ respectively. 
It is augmented with a conflicting clause $a_1 \lor \obar{b}_2 \lor \obar{c}_4$ which makes the formula unsatisfiable.
Our lemma will be the derivation of units $\obar{z}_1, z_2$, and $z_4$ as in Figure~\ref{fig:formula:example}.
Once proven, we can then apply this lemma to the variables in each block, with the applications to blocks $b$ and $c$ under qualification,
to derive units $\obar{a}_1$, $b_2$, and $c_4$, obtaining a contradiction.
\begin{description}
    \item[Prefixes] The first step in the embedding procedure adds prefix clauses. 
        Since there are $K=3$ applications, we need $K-1=2$ sets of prefix variables,
        each for deriving $m=3$ consequent clauses. To do this, the procedure adds clauses: 
\begin{center}
    \setlength{\extrarowheight}{1.25pt}
\begin{tabular}[t]{@{}>{\centering\arraybackslash}p{82.3pt}@{}}
\toprule
$a$'s prefix chain \\
\midrule
\begin{tabular}{@{}r@{\ $\leftrightarrow$\ }l@{}}
  $s_1$ & $\obar{a}_1$ \\
  $s_2$ & $s_1 \land a_2$ \\
  $s_3$ & $s_2 \land a_4$
\end{tabular} \\
\bottomrule
\end{tabular}
\quad
\begin{tabular}[t]{@{}>{\centering\arraybackslash}p{82.3pt}@{}}
\toprule
$b$'s prefix chain \\
\midrule
\begin{tabular}{@{}r@{\ $\leftrightarrow$\ }l@{}}
  $s_4$ & $s_3 \land (\obar{b}_1 \lor \obar{a}_2)$ \\
  $s_5$ & $s_4 \land (b_2 \lor \obar{a}_2)$ \\
  $s_6$ & $s_5 \land (b_4 \lor \obar{a}_2)$
\end{tabular} \\
\bottomrule
\end{tabular}
\end{center}
        Variable $s_r$ is equivalent to the conjunction of the first $r$ instantiated consequent clauses. This gives the procedure access to ``selectors''
        for application $k$ of the form $\obar{s}_{km} \land s_{(k-1)m}$ which intuitively mean ``we know the consequents
        of instantiations through $k-1$ are valid, but we don't know about instantiation $k$.''
    \item[Bindings] The key insight of this construction is the introduction of bindings conditional on selectors. 
        Throughout the proof, we want to maintain the invariant that lemma variables $\zdata$
        are bound to the earliest application that has not been checked. To do this, step $2$ binds
        the parameters conditionally on the selectors.
        \begin{center}

    \setlength{\extrarowheight}{1.25pt}
            \begin{tabular}[t]{@{}>{\centering\arraybackslash}p{78pt}@{}}
        \toprule
        $a$'s bindings \\
        \midrule
            $\obar{s}_3 \rightarrow z_1 \leftrightarrow a_1$ \\
            $\obar{s}_3 \rightarrow z_2 \leftrightarrow a_2$ \\
            $\obar{s}_3 \rightarrow z_3 \leftrightarrow a_3$ \\
            $\obar{s}_3 \rightarrow z_4 \leftrightarrow a_4$ \\
        \bottomrule
        \end{tabular}
        \quad
            \begin{tabular}[t]{@{}>{\centering\arraybackslash}p{78pt}@{}}
        \toprule
        $b$'s bindings \\
        \midrule
            $\obar{s}_6 \land s_3 \rightarrow z_1 \leftrightarrow b_1$ \\
            $\obar{s}_6 \land s_3 \rightarrow z_2 \leftrightarrow b_2$ \\
            $\obar{s}_6 \land s_3 \rightarrow z_3 \leftrightarrow b_3$ \\
            $\obar{s}_6 \land s_3 \rightarrow z_4 \leftrightarrow b_4$ \\
        \bottomrule
        \end{tabular}
        \quad
            \begin{tabular}[t]{@{}>{\centering\arraybackslash}p{78pt}@{}}
        \toprule
        $c$'s bindings \\
        \midrule
            $s_6 \rightarrow z_1 \leftrightarrow c_1$ \\
            $s_6 \rightarrow z_2 \leftrightarrow c_2$ \\
            $s_6 \rightarrow z_3 \leftrightarrow c_3$ \\
            $s_6 \rightarrow z_4 \leftrightarrow c_4$ \\
        \bottomrule
        \end{tabular}
        \end{center}
    \item[Antecedent] Next, the procedure adds the lemma's antecedent clauses. Since the previous step introduced the bindings,
        the antecedent clauses are neither RUP nor RAT, in general. Thus, step $3$ adds guarded versions that can be activated
        later.         
        \begin{center}
        \vspace{-10pt}
    \setlength{\extrarowheight}{1.25pt}
        \begin{tabular}[t]{c}
        \toprule
        Guarded Antecedent \\
        \midrule
            $t \rightarrow z_1\lor z_2\lor\zneg_3$ \\
            $t \rightarrow \zneg_1\lor\zneg_2\lor z_3$ \\
            $t \rightarrow z_2\lor z_3\lor\zneg_4$ \\
            $t \rightarrow \zneg_2\lor\zneg_3\lor z_4$ \\
            $t \rightarrow z_1\lor z_3\lor z_4$ \\
            $t \rightarrow \zneg_1\lor\zneg_3\lor\zneg_4$ \\
            $t \rightarrow \zneg_1\lor z_2\lor z_4$ \\
        \bottomrule
        \end{tabular}
        \quad
        \begin{tabular}[t]{c}
        \toprule
        Activators\\
        \midrule
            $\phantom{a_2 \land s_3 \land\,} \obar{s}_3 \rightarrow t$\\
            ${a_2 \land s_3 \land\,} \obar{s}_6 \rightarrow t$\\
            $\phantom{a_2 \land\,} \obar{b}_1 \land s_6 \rightarrow t$\\
            ~\\
            ~\\
            ~\\
            ~\\
        \bottomrule
        \end{tabular}
        \end{center}
    \item[Prove Once] Step $4$ proves the lemma on the $Z$ variables. It simply replays the proof, guarded by $t$ as follows:
            \vspace{-5pt}
        \begin{center}

    \setlength{\extrarowheight}{1.25pt}
        \begin{tabular}{r@{\ $\rightarrow$\ }l}
        \toprule
        \multicolumn{2}{c}{Proof Replay} \\
        \midrule
        $t$        & $z_2 \lor z_4$ \\
        $t$        & $z_2$ \\
        delete $t$ & $z_2 \lor z_4$ \\
        $t$        & $z_4$ \\
        $t$        & $\obar{z}_1$ \\
        \bottomrule
        \end{tabular}

        \end{center}
    \item[Prefix Units] 
        At this point, the prefix units follow in order, $s_1, \ldots, s_6$. 
    \item[Consequent] All that is left to do is derive the consequent clauses by reading them off:
        \begin{center}
    \setlength{\extrarowheight}{1.25pt}
            \begin{tabular}[t]{c}
            \toprule
            $a$'s consequent \\
            \midrule
                $\obar{a}_1$\\
                $a_2$\\
                $a_4$\\
            \bottomrule
            \end{tabular}
            \quad
            \begin{tabular}[t]{c}
            \toprule
            $b$'s consequent \\
            \midrule
                $\obar{b}_1 \lor \obar{a}_2$\\
                $b_2\lor \obar{a}_2$\\
                $b_4 \lor \obar{a}_2$\\
            \bottomrule
            \end{tabular}
            \quad
            \begin{tabular}[t]{c}
            \toprule
            $c$'s consequent\\
            \midrule
                $\obar{c}_1 \lor b_1$\\
                $c_2 \lor b_1$\\
                $c_4 \lor b_1$\\
            \bottomrule
            \end{tabular}
        \end{center}
        With the consequents derived, the empty clause is RUP, showing that the formula is unsatisfiable.
\end{description}
%
\section{The DRATL Format for Lemma-Based Proofs}

\begin{figure}[tb]
\newcommand{\nt}[1]{\mathit{#1}}   
\newcommand{\tm}[1]{\texttt{#1}}   
\centering
$\begin{array}{r@{\;\;}c@{\;\;}l@{\qquad}l}
\nt{proof}  & ::=  & \nt{step}^{*} \\[3pt]
\nt{step}   & ::=  & \nt{clause}                 & \text{add a clause (\drat)} \\
            & \mid & \tm{d}\ \nt{clause}         & \text{delete a clause (\drat)} \\
            & \mid & \nt{lemma}                  & \text{declare and prove a lemma} \\
            & \mid & \nt{apply}                  & \text{apply a lemma} \\[3pt]
\nt{lemma}  & ::=  & \tm{L}\ \nt{id}\ n\ x\ a\ m & \text{$n$ parameters, $x$ auxiliary variables} \\
            &      & \nt{clause}^{a}             & \text{$a$ antecedent clauses, over variables $1,\ldots,n$} \\
            &      & \nt{clause}^{m}             & \text{$m$ consequent clauses, over variables $1,\ldots,n$} \\
            &      & \nt{step}^{*}               & \text{proof of the lemma, over variables $1, \ldots, n +x$} \\
            &      & \tm{e}\ \nt{id} \\[3pt]
\nt{apply}  & ::=  & \tm{l}\ \nt{id}\ \nt{lit}^{n}\ \tm{0}\ [\,\nt{lit}^{*}\ \tm{0}\,] & \text{arguments, optional qualifier} \\[3pt]
\nt{clause} & ::=  & \nt{lit}^{*}\ \tm{0}
\end{array}$
    \caption{Grammar for the DRATL format}
    \label{fig:grammar}
\end{figure}
\begin{figure}[tb]
\centering{
\setlength{\tabcolsep}{6pt}
\begin{tabular}{cll}
\toprule
Action & Clause & Comment \\
\midrule
L & $1\;\;4\;\;0\;\;7\;\;3$ & $4$ parameters, $0$ auxiliaries, $7$ antecedents, $3$ consequents \\
  & $z_1\lor z_2\lor\zneg_3$ & antecedent \\
  & $\zneg_1\lor\zneg_2\lor z_3$ & \\
  & $z_2\lor z_3\lor\zneg_4$ & \\
  & $\zneg_2\lor\zneg_3\lor z_4$ & \\
  & $z_1\lor z_3\lor z_4$ & \\
  & $\zneg_1\lor\zneg_3\lor\zneg_4$ & \\
  & $\zneg_1\lor z_2\lor z_4$ & \\
  & $\zneg_1$ & consequent\\
  & $z_2$ & \\
  & $z_4$ & \\
a & $z_2\lor z_4$ & proof of the consequent \\
a & $z_2$ & \\
d & $z_2 \lor z_4$\\
a & $z_4$ & \\
a & $\zneg_1$ & \\
e & $1$ & end of lemma $1$ \\
l & $1 : a_1, a_2, a_3, a_4$ & consequent $\obar{a}_1$, $a_2$, $a_4$ \\
l & $1 : b_1, b_2, b_3, b_4\;\;[\,\obar{a}_2\,]$ & consequent $\obar{b}_1\lor\obar{a}_2$, $b_2\lor\obar{a}_2$, $b_4\lor\obar{a}_2$ \\
l & $1 : c_1, c_2, c_3, c_4\;\;[\,b_1\,]$ & consequent $\obar{c}_1\lor b_1$, $c_2\lor b_1$, $c_4\lor b_1$ \\
a & $\bot$ & RUP from the consequents and the conflicting clause \\
\bottomrule
\end{tabular}
} 
\caption{The DRATL proof of the formula in Figure~\ref{fig:3sat} with the lemma of Figure~\ref{fig:formula:example}.
  A qualifier is shown in brackets after the arguments. The presentation is stylized, using literal symbols instead of numbers.}
\label{fig:dratl-example}
\end{figure}

DRATL extends DRAT to enable proofs structured as
hierarchies of lemmas.
DRATL adds
two extensions to the DRAT format: lemma definitions and lemma instantiations. The full grammar can be seen in Figure~\ref{fig:grammar}.
An example DRATL proof, replicating the one from Section~\ref{sec:embedding}, is shown in Figure~\ref{fig:dratl-example}. The new components are as follows:
\begin{description}
    \item[Lemma Definition] A lemma is defined by first giving a header with a user-specified, unique numeric ID,  
        as well as the number of parameters, auxiliary variables, antecedent clauses, and consequent clauses.
        Then the antecedent and consequent clauses are listed, followed by the proof of the consequent from the antecedent. 
    \item [Lemma Application]
    An application consists of a lemma ID and one literal per parameter, where the literal in position $i$
    specifies the value of $\sigma(z_i)$.
    The instantiation concludes with an optional list of literals forming the qualifying literals $Q$.
\end{description}
\subsection{Elaboration}
Our elaborator translates a DRATL proof into a DRAT proof following a variation of the procedure
in Figure~\ref{fig:compact}. All of the additional complexity arises from nested lemmas, where a lemma $i$ is instantiated as part of
the proof of another lemma $j$. In this section, we define what it means for a DRATL proof to be semantically valid,
describe how we elaborate such files, and prove that the elaboration results in a valid DRAT proof.

A \textit{scope} is defined to be either the top-level proof or the proof of a lemma. 
A scope's formula is the set of clauses relative to which the proof is checked. At the top level, it is the original formula $F$
and within a lemma it is the antecedent. A scope's proof and consequent are defined similarly.
A scope's proof can be divided into two parts: the \textit{preamble},
consisting of all steps
before any lemma is applied, and the \textit{body}, consisting of the remainder of the proof steps.
For a DRATL proof to be valid, we require the following:
\begin{enumerate}
    \item For each lemma application, the conditions described in Figure~\ref{fig:compact} are met with respect to its enclosing scope's formula and preamble.
    \item If lemma $i$ is used in the proof of lemma $j$, lemma $j$ is not used, directly or indirectly, in the proof of lemma $i$.
    \item In all scopes, the RAT steps in the proof's body must be valid with respect to all instantiated consequents of all lemma applications in the scope.
    \item Deletion steps are not allowed to delete instantiated consequent clauses.
    \item Every scope's proof is a valid derivation of its consequent from its formula when lemma applications, given by
        instantiation $\sigma$ and qualifier $Q$, are interpreted as the instantiated consequent $\sigma(C) \lor Q$
        for $C \in \CC$.
\end{enumerate}
Item $3$ is required because DRAT is non-monotonic. While it appears restrictive, it did not hinder
our ability to write compact DRATL proofs. Specifically, we found that only using RAT steps in a proof's preamble was
sufficiently expressive. Item $4$ is required because lifting all consequent clauses to the top exposes them to premature deletion.

Given a valid DRATL proof, the elaboration algorithm follows a simple procedure. First, the elaborator emits the top level's preamble.
When the first lemma application is encountered, we topologically sort the lemmas by application. That is,
if lemma $i$ is used in the proof of lemma $j$, then $i$ comes before $j$ in this ordering. Then, it applies the steps of
Figure~\ref{fig:compact} in two phases. The elaborator's setup phase applies steps $1$ through the preambles of step $4$ on each lemma in reverse topological
order. Then the conclusion
phase applies the bodies of the proofs in step $4$, as well as steps $5$ through $7$, on each lemma in topological order.
Moreover, when a lemma $i$ is used in the proof of lemma $j$, it is implicitly qualified by $\obar{t}_j$, the
negated guard literal for lemma $j$. Finally, the body of the top level's proof is replayed.

This construction uses two phases to satisfy opposing dependency orders; a single one does not suffice. If lemma $i$ is used in the proof of lemma $j$,
then an instantiation of lemma $i$, namely the one which is used in lemma $j$, has antecedent clauses that
are RUP from the antecedent and preamble of lemma $j$. This would seem to require that lemma $j$  be elaborated before lemma $i$. 
On the other hand, the consequent for lemma $j$ depends on the instantiated consequent from lemma $i$. 
This would seem to require that lemma $i$ be elaborated before lemma $j$. The two-phase approach resolves this tension.
We now prove the correctness of this approach.

\begin{theorem}[Elaboration Correctness]    The elaboration of a valid DRATL proof results in a valid DRAT proof.
\end{theorem}
\begin{proof}
    To show that the nested elaboration algorithm results in a valid DRAT proof, we argue that each of the
    justifications from Figure~\ref{fig:compact} still applies.

    Fix a lemma $i$ which may be applied within lemma $j$.
    Since the prefix, problem, auxiliary, and $t$ variables are disjoint for each lemma,
    steps $1$ and $2$ work as they did in Theorem~\ref{thm:embedding}. The only complexity in step $3$ is due to
    the RAT check on $t$ possibly depending on the antecedent and preamble for lemma $j$. By the reverse topological ordering of the setup phase, the
    antecedent and preamble for lemma $j$ have already been emitted, but their clauses are guarded by $t_j$.     
    However, during elaboration, the application of lemma $i$ inside lemma $j$ is implicitly qualified by $\obar{t}_j$, allowing the argument to follow exactly as in Theorem~\ref{thm:embedding}.
    The last step of the setup phase, the preamble, has no lemma instantiations, and is the same as before.

    In the conclusion phase of lemma $i$, the only clauses remaining that mention the variables of $i$
    are from its own elaboration and the instantiated consequents of the lemmas applied in its proof. This is because 
    the topological ordering of the conclusion phase guarantees that all lemmas used in the proof of lemma $i$
    have already been elaborated and undergone cleanup,
    leaving only their instantiated consequents.
    Due to the restriction on deleting instantiated consequents, 
    every clause available at a given step of the DRATL proof for lemma $i$ is also available,
    in guarded form,
    at the corresponding step of the elaborated proof. 
    In this manner, RUP steps remain valid by monotonicity, while RAT steps remain valid by Item~3, 
    allowing the body to be replayed.
    Steps $5$ through $7$ involve only the clauses of lemma $i$ and are therefore unchanged.

    After all lemmas have been elaborated, at the top level, the remaining proof body is valid with respect to all
    of the instantiated consequents and is otherwise a valid DRAT proof. Therefore, the resulting elaborated
    DRAT proof is valid.
\end{proof}

\section{Experimental Results}
\label{sec:results}

Three case studies probe the approach from different sides.  Using
lemmas for the mutilated chessboard enables an effective strategy
based on extended resolution to scale even further by reducing the
overhead of adding formulas implementing modular arithmetic.
On the queen graphs, a single lemma enables reducing many copies of a subformula describing a clique.
Lemmas for the Mycielski graph follow its inductive structure, enabling a factorial reduction in size.

\newcommand{\kissat}{\textsc{kissat}}
\newcommand{\pgbdd}{\textsc{pgbdd}}
\newcommand{\pgpbs}{\textsc{pgpbs}}

\definecolor{redorange}{rgb}{0.878431, 0.235294, 0.192157}
\definecolor{lightblue}{rgb}{0.95, 0.9, 0.99}
\definecolor{clearyellow}{rgb}{0.964706, 0.745098, 0}
\definecolor{clearorange}{rgb}{0.917647, 0.462745, 0}
\definecolor{mildgray}{rgb}{0.54902, 0.509804, 0.47451}
\definecolor{softblue}{rgb}{0.643137, 0.858824, 0.909804}
\definecolor{bluegray}{rgb}{0.141176, 0.313725, 0.603922}
\definecolor{lightgreen}{rgb}{0.9, 0.99, 0.9}
\definecolor{redpurple}{rgb}{0.835294, 0, 0.196078}
\definecolor{midblue}{rgb}{0, 0.592157, 0.662745}
\definecolor{clearpurple}{rgb}{0.67451, 0.0784314, 0.352941}
\definecolor{browngreen}{rgb}{0.333333, 0.313725, 0.145098}
\definecolor{darkestpurple}{rgb}{0.396078, 0.113725, 0.196078}
\definecolor{greypurple}{rgb}{0.294118, 0.219608, 0.298039}
\definecolor{darkturquoise}{rgb}{0, 0.239216, 0.298039}
\definecolor{darkbrown}{rgb}{0.305882, 0.211765, 0.160784}
\definecolor{midgreen}{rgb}{0.560784, 0.6, 0.243137}
\definecolor{darkred}{rgb}{0.576471, 0.152941, 0.172549}
\definecolor{darkpurple}{rgb}{0.313725, 0.027451, 0.470588}
\definecolor{darkestblue}{rgb}{0, 0.156863, 0.333333}
\definecolor{lightpurple}{rgb}{0.776471, 0.690196, 0.737255}
\definecolor{softgreen}{rgb}{0.733333, 0.772549, 0.572549}
\definecolor{medgreen}{rgb}{0.34, 0.65, 0.34}
\definecolor{offwhite}{rgb}{0.839216, 0.823529, 0.768627}

\subsection{Mutilated Chessboard}
\label{sect:mchess}

\begin{figure}[tb]
\centering
\begin{tikzpicture}[scale = 1.0]
          \begin{axis}[mark options={scale=0.8},grid=both, grid style={black!10}, xmode=log, ymode=log, legend style={at={(0.99,0.58)}}, legend cell align={left},
                              x post scale=1.55, y post scale=1.3,
                              xlabel=$n$, xmin=10,xmax=4000, xtick={10,100,1000,10000}, xticklabels={10,100,1000,{10,000}},
                              ylabel=Clauses, ymin=100,ymax=100000000, 
            ]
\addplot [color=clearpurple, mark=o] coordinates { (4.0,58.0) (6.0,204.0) (8.0,942.0) (10.0,9180.0) (12.0,48296.0) (14.0,691551.0) (16.0,4503635.0) (18.0,17281161.0) (20.0,65376268.0) (22.0,136883320.0)};
\addplot [color=bluegray, mark=triangle] coordinates { (20.0,7598.0) (30.0,22879.0) (40.0,48967.0) (50.0,91665.0)};
\addplot [color=redorange, mark=triangle] coordinates { (20.0,501766.0) (30.0,2489657.0) (40.0,7776380.0) (50.0,18845988.0)};
\addplot [color=darkestblue, mark=square] coordinates { (4.0,1493.0) (8.0,12120.0) (12.0,34830.0) (16.0,72573.0) (20.0,128268.0) (24.0,204895.0) (28.0,305414.0) (32.0,432785.0) (36.0,589968.0) (40.0,779923.0) (44.0,1005610.0) (48.0,1269989.0) (52.0,1576020.0) (56.0,1926663.0) (60.0,2324878.0) (64.0,2773625.0) (68.0,3275864.0) (72.0,3834555.0) (76.0,4452658.0) (80.0,5133133.0) (84.0,5878940.0) (88.0,6693039.0) (92.0,7578390.0) (96.0,8537953.0) (100.0,9574688.0) (104.0,10691555.0) (108.0,11891514.0) (112.0,13177525.0) (116.0,14552548.0) (120.0,16019543.0) (124.0,17581470.0) (128.0,19241287.0)};
\addplot [color=midblue, mark=diamond] coordinates { (4.0,1413.0) (8.0,13153.0) (12.0,40298.0) (16.0,76923.0) (20.0,129472.0) (24.0,203562.0) (28.0,288249.0) (32.0,381243.0) (40.0,630087.0) (48.0,916145.0) (56.0,1269555.0) (64.0,1668356.0) (80.0,2699327.0) (96.0,3944206.0) (112.0,5470256.0) (128.0,7152924.0)};
\addplot [color=darkred, mark=diamond*] coordinates { (6.0,1421.0) (10.0,4437.0) (14.0,9117.0) (18.0,15461.0) (24.0,28097.0) (32.0,50769.0) (42.0,88469.0) (56.0,158721.0) (74.0,278997.0) (100.0,512217.0) (140.0,1008297.0) (180.0,1670777.0) (240.0,2976497.0) (320.0,5299857.0) (420.0,9140057.0) (560.0,16263537.0) (740.0,28417497.0) (1000.0,51922017.0) (1300.0,87778617.0)};
\addplot [color=midgreen, mark=triangle*] coordinates { (6.0,1534.0) (10.0,3416.0) (14.0,5954.0) (18.0,8774.0) (24.0,15169.0) (32.0,22195.0) (42.0,32924.0) (56.0,51687.0) (74.0,80988.0) (100.0,148539.0) (140.0,241093.0) (180.0,384489.0) (240.0,574495.0) (320.0,1055701.0) (420.0,1533521.0) (560.0,2448197.0) (740.0,4012207.0) (1000.0,7353598.0) (1400.0,13010586.0) (1800.0,20422150.0) (2400.0,34212832.0) (3200.0,57998435.0)};
\addplot [color=darkturquoise, mark=square*, mark options={scale=0.7, opacity=0.5}] coordinates { (6.0,2729.0) (10.0,4745.0) (14.0,7154.0) (18.0,9689.0) (24.0,16967.0) (32.0,20747.0) (42.0,29440.0) (56.0,45419.0) (74.0,67247.0) (100.0,108364.0) (140.0,213336.0) (180.0,300020.0) (240.0,452867.0) (320.0,792214.0) (420.0,1166826.0) (560.0,2024129.0) (740.0,3257893.0) (1000.0,5737605.0) (1400.0,10487947.0) (1800.0,16380163.0) (2400.0,28672933.0) (3200.0,48765634.0)};
\addplot [color=clearorange, mark=*] coordinates { (6.0,184.0) (10.0,584.0) (14.0,1208.0) (18.0,2056.0) (24.0,3748.0) (32.0,6788.0) (42.0,11848.0) (56.0,21284.0) (74.0,37448.0) (100.0,68804.0) (140.0,135524.0) (180.0,224644.0) (240.0,400324.0) (320.0,712964.0) (420.0,1229764.0) (560.0,2188484.0) (740.0,3824324.0) (1000.0,6988004.0) (1400.0,13703204.0) (1800.0,22658404.0) (2400.0,40291204.0) (3200.0,71641604.0)};
            \legend{
              \scriptsize \textsf{\kissat},
              \scriptsize \textsf{PR proof~\cite{heule:nfm:2019}},
              \scriptsize \textsf{PR converted to DRAT~\cite{heule:nfm:2019}},
              \scriptsize \textsf{BDD, Column scan~\cite{bryant:tocl:2023}},
              \scriptsize \textsf{BDD, Mod-3 equations~\cite{bryant:tacas:2022}},
              \scriptsize \textsf{Mod-3 propagation, Flat},
              \scriptsize \textsf{Mod-3 propagation, Strips},
              \scriptsize \textsf{Mod-3 propagation, Tiled},
              \scriptsize \textsf{Input formula},
            }
\addplot [color=midgreen, mark=triangle*] coordinates { (6.0,1534.0) (10.0,3416.0) (14.0,5954.0) (18.0,8774.0) (24.0,15169.0) (32.0,22195.0) (42.0,32924.0) (56.0,51687.0) (74.0,80988.0) (100.0,148539.0) (140.0,241093.0) (180.0,384489.0) (240.0,574495.0) (320.0,1055701.0) (420.0,1533521.0) (560.0,2448197.0) (740.0,4012207.0) (1000.0,7353598.0) (1400.0,13010586.0) (1800.0,20422150.0) (2400.0,34212832.0) (3200.0,57998435.0)};
\addplot [color=darkturquoise, mark=square*, mark options={scale=0.7, opacity=0.5}] coordinates { (6.0,2729.0) (10.0,4745.0) (14.0,7154.0) (18.0,9689.0) (24.0,16967.0) (32.0,20747.0) (42.0,29440.0) (56.0,45419.0) (74.0,67247.0) (100.0,108364.0) (140.0,213336.0) (180.0,300020.0) (240.0,452867.0) (320.0,792214.0) (420.0,1166826.0) (560.0,2024129.0) (740.0,3257893.0) (1000.0,5737605.0) (1400.0,10487947.0) (1800.0,16380163.0) (2400.0,28672933.0) (3200.0,48765634.0)};
          \end{axis}
\end{tikzpicture}

\caption{Total number of clauses in formula and proofs of $n \times n$
  mutilated chessboard problems.  Historical methods were limited to
  $n \leq 128$.  Direct generation with mod-3 propagation scales beyond
  $n=1000$.  Using lemmas (strips or tiled) greatly decreases the
  proof size for large values of $n$, even becoming more compact than the formulas themselves.}
\label{fig:data:mchess}
\end{figure}
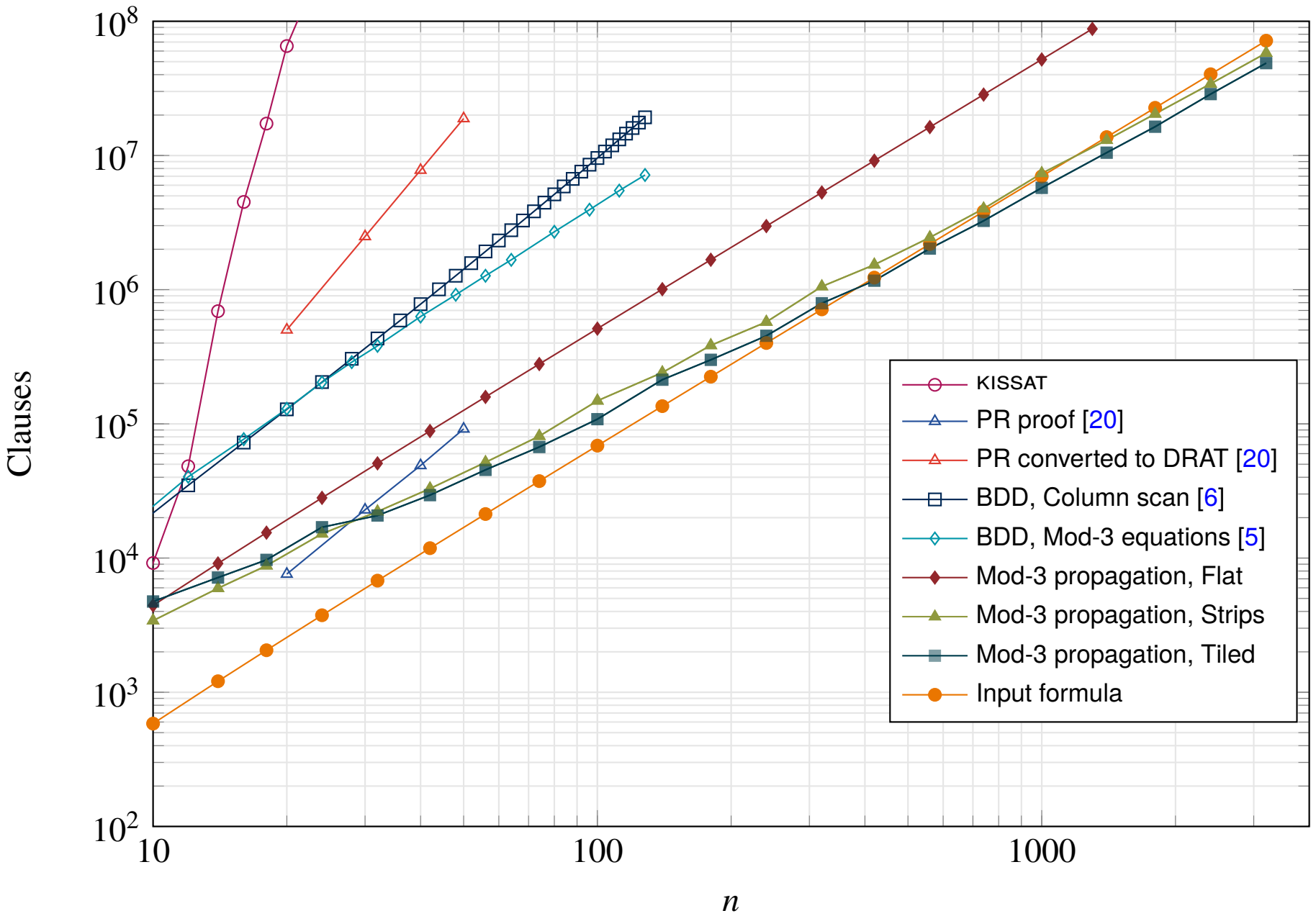  

The mutilated chessboard problem has long served as a challenge for
automated reasoning tools~\cite{mccarthy:memo:1964}.  The problem
considers an $n \times n$ chessboard, with the corners on the upper
left and the lower right removed.  It attempts to tile the board with
dominoes, with each domino covering two squares.  For even values of $n$, both removed
squares are white.  Since each domino covers one white and one
black square, no tiling is possible.

Labeling the squares of a chessboard as having rows 1 (upper) to $n$
(lower) and columns 1 (left) to $n$ (right), the standard Boolean
encoding of the problem introduces two sets of variables describing
the placement of horizontal and vertical dominoes, respectively.
Variable $h_{r,c}$ for $1 \leq r \leq n$ and $1 \leq c < n$ encodes
whether a domino is placed horizontally on row $r$ spanning columns
$c$ and $c+1$.  Variable $v_{r,c}$ for $1 \leq r < n$ and
$1 \leq c \leq n$ encodes whether a domino is placed vertically on column $c$
spanning rows $r$ and $r+1$.  The upper-left and lower-right corners
are omitted via unit clauses $\obar{h}_{1,1}$, $\obar{v}_{1,1}$, $\obar{h}_{n,n-1}$, and $\obar{v}_{n-1,n}$.
For every other square, there are 2--4 variables
describing possible placements of dominoes on that square.  For each square, 2--7 clauses
constrain
exactly one
of these to be set to $1$.  
It can be shown that any
resolution proof of unsatisfiability for this encoding must be of exponential
size~\cite{alekhnovich:tcs:2004}.  The plot labeled \kissat{} in
Figure~\ref{fig:data:mchess} demonstrates how difficult this problem
is for CDCL solvers, showing the number of proof clauses as a function
of $n$.  Even for $n=20$, \kissat{} generates a proof with over 65~million clauses.

Others have found ways to overcome exponential performance on this
problem, such as by applying symmetry reduction via the
\emph{propagation-redundancy} (PR) proof system~\cite{heule:nfm:2019}.
The PR proofs can then be translated into DRAT, with polynomial but significant
overhead.  The plots in Figure~\ref{fig:data:mchess}
labeled ``PR proof'' and ``PR converted to DRAT'' show their published results, reaching $n=50$.
Although it can be seen that the
PR proofs, and their translation to DRAT, greatly improve on
\kissat{}, other approaches generally outperform the DRAT versions,
and the results reported here even improve on the PR proof sizes.

More recent work using proof-generating binary decision diagram (BDD)
packages demonstrated success for up to $n=128$ using two different
approaches.  One used a strategy inspired by symbolic model checking
to scan from the left side of the board to the right, tracking the
possible sets of assignments to the variables $h_{r,c}$ for $1 \leq r
\leq n$ as $c$ ranges from $1$ to $n-1$~\cite{bryant:tocl:2023}.
Their published results are shown as the plot ``BDD, Column scan.''
The second converted the problem into one involving pseudo-Boolean
constraints, encoding the mismatch between the total number of white
and black squares.  They observed that the arithmetic could be computed
modulo-3 and still capture the mismatch~\cite{bryant:tacas:2022}.  Their results are shown in the plot
``BDD, Mod-3 equations.''

Rather than using BDDs as an intermediate step, we can combine the
ideas of column scanning and mod-3 arithmetic directly into an
extended resolution proof.  That is, we set up formulas encoding mod-3
counters summing $w_{r,c} \cdot h_{r,c}$ for $1 \leq r \leq n$, where
$w_{r,c} = (r+c \bmod 2) + 1$.  For $c = 1$, the sum will be 2.  Each
successive column will preserve the sum, and then a conflict will
arise at the final column.  The plot labeled ``Mod-3 propagation,
Flat'' shows that this approach scales far better than any previous
method.  Whereas all the previously cited methods have DRAT proofs
exceeding one million clauses for $n \geq 50$, the direct encoding
reaches $n=140$ before hitting that limit.

Lemmas allow us to push the scaling of modular arithmetic even
further, by reducing the number of clauses added to encode and reason about mod-3
counters.  For a width $w$ where $n-2$ is a multiple of $w$, we can
divide the board into the left column, a series of strips $w$ columns
wide, and the right column.  A single lemma suffices to show that the
count at the left boundary of a strip is preserved when scanning to
its right boundary.  By setting $w$ to around $\sqrt{n}$, we get
balance between the number of mod-3 counters that connect the
successive strips and the number of counters within the lemma between
the columns of a single strip.  When there are no suitable divisors
of $n-2$, we may need to add a second lemma for a width $w' < w$.
These results are shown as the plot ``Mod-3 propagation, Strips'' in
Figure~\ref{fig:data:mchess}.  For large values of $n$, we can see
that the use of lemmas substantially reduces the proof size.  These
proofs scale to $n=318$ before reaching one million clauses.  In
keeping with the DRATL requirements, all of the mod-3 counters must be
introduced as part of the preamble for the top-level proof.
Remarkably, for large values of $n$, these proofs become smaller than the input formula, with the crossover occurring around $n=1100$.

As a final optimization, we can subdivide the strips of width $w$ into
columns of height $h$ and encode summations of both sets of variables
$h_{r,c}$ and $v_{r,c}$.  As before, we may
need to add an extra lemma for a smaller width, as well as lemmas for
smaller heights.  The exact setting of the parameters requires
balancing many considerations, but we were able to train a regression
model to select the parameters.  These results are shown as the plot
``Mod-3 propagation, Tiled'' in Figure~\ref{fig:data:mchess}.  As can
be seen, it improves on the strips approach only for large values of
$n$, reaching one million clauses with $n=384$.  For $n \geq 380$, these proofs are smaller than the original input formula.

As these results show,  mod-3 propagation alone enables compact proofs of
the mutilated chessboard problem, but the use of lemmas greatly
reduces the amount of overhead.  At $n=1000$, the
tiled proofs are over $9\times$ smaller than the flat ones.

The scaling of these proofs presents a challenge for checking with our
verification toolchain, consisting of
\textsc{drat-trim}~\cite{heule:arxiv:2016}, a tool for expanding a DRAT proof into a more detailed form,
followed by the formally verified
\textsc{cake\_lpr} checker~\cite{tan:sttt:2023}.
We can only validate the flat proofs for up to $n=320$ and
the strips and tiled proofs for up to $n=420$.  The runtimes for
\textsc{drat-trim} become impractical beyond that
point, exceeding 30,000 seconds.  
Enabling
more efficient checking is a clear priority for future work.

\subsection{Queen Graph Coloring}

The queen graph $Q_n$ has the squares of an $n\times n$ board as
vertices, adjacent when a queen moves between them. $Q_n$ is not
$n$-colorable for $n \in \{8,9,10\}$. The
$n=10$ formula, even when extended with symmetry breaking (fixing the diagonal), is hard for CDCL solvers. 
Solvers cannot easily infer surjectivity on the maximum cliques: in an $n$-coloring,
every row, column, and long diagonal uses every color. This is a
pigeonhole-style consequence with no short resolution proof. Deriving
these \emph{clique hints}
$H_{Q,k}=\bigvee_{v\in Q} x_{v,k}$ for clique $Q$ and color $k$
collapses the instances. For $n \geq 11$, $Q_n$ is conjectured to be
$n$-colorable. Colorings are known for $11 \leq n \leq 26$~\cite{chvatal:queengraphs}. Clique hints also
boost performance on satisfiable formulas. 

We derive the clique hints using a single lemma with $n^2$ parameters
representing the assignment of $n$ colors to $n$ vertices. Its antecedent clauses
require each vertex to have at least one color and prohibit two vertices from
sharing a color. Its $n$ consequent clauses state that every color occurs in the clique.
The lemma uses the chain derivation of Grosof, Zhang, and Heule~\cite{grosof:pos:2022},
with $\tfrac52 n^3$ steps and all newly introduced variables treated as auxiliaries.
We apply the lemma to each of the $2n+2$ maximum cliques: the $n$ rows, $n$ columns,
and two long diagonals. In each application, the instantiated antecedent consists of input
clauses. Deriving all $n$ hints together allows them to share the same parameter
bindings. An alternative lemma deriving a single clique hint would also require $n^2$ parameters.
Applying it separately for each color would increase the number of binding clauses by a factor of $n$.

The pipeline verifies \emph{end to end} from the plain formula,
within the substitution redundancy (SR) proof system~\cite{buss:sat:2019}, a generalization of DRAT.
First, for symmetry breaking, $n^2$ SR steps fix the main diagonal to $n$ different colors,
each witnessed by a color swap in the style of orbitopal fixing~\cite{anders:sat:2026}.
Second, the clique hints are derived via our lemma approach and turned into DRAT using the elaborator.
Finally, {\tt CaDiCaL} (version 3.0) produces that last part of the proof by
solving the formula extended with the symmetry-breaking unit clauses and the clique hints.
The concatenation is a single DSR proof of the original formula,
checked by \texttt{dsr-trim}, whose annotated output is then checked
by \texttt{srcheck}, the formally verified checker
of~\cite{codel:fmcad:2024}, in under a dozen seconds per instance. With symmetry breaking, but without the hints,
{\tt CaDiCaL} takes $5.7$ seconds at $n=8$ and $801$ seconds at $n=9$, and
does not solve $n=10$. With symmetry breaking and clique hints, all three instances are refuted in
seconds (Table~\ref{tab:queens}).

\begin{table}[t]
\caption{Queen graphs: one hint lemma per size, proved once and
applied to all $2n{+}2$ maximum cliques. The first three columns show
the size of the board and the number of clauses + symmetry-breaking units and the CDCL runtime without clique hints, respectively. 
The fourth column shows the number of lemma applications, 
followed by the number of added clique hints (consequent clauses). 
The last four columns show the size of deriving every hint by
replaying the derivation once per hint (no lemmas), the sizes of the
elaborated lemma proof and of the end-to-end proof, and the time to
check the latter. 
The $n=10$ instance was not solved within an hour.}\label{tab:queens}
\centering
\setlength{\tabcolsep}{3.4pt}
\begin{tabular}{rrrccrrcrr}
\toprule
 & \multicolumn{1}{c}{clauses +} & \multicolumn{1}{c}{time w/o} & \multicolumn{1}{c}{lemma} & \multicolumn{1}{c}{clique} & \multicolumn{1}{c}{time w/} & \multicolumn{1}{c}{flat} & \multicolumn{1}{c}{} & \multicolumn{1}{c}{\!end-to-end\!} & \multicolumn{1}{c}{} \\
\multicolumn{1}{c}{$n$} & \multicolumn{1}{c}{sym.\ units} & \multicolumn{1}{c}{hints} & \multicolumn{1}{c}{appl.} & \multicolumn{1}{c}{hints} & \multicolumn{1}{c}{hints} & \multicolumn{1}{c}{total} & \multicolumn{1}{c}{\!elaborated\!} & \multicolumn{1}{c}{proof} & \multicolumn{1}{c}{check} \\
\midrule
$8$  &    $7{,}680 + 8\phantom{0}$ &      $5.7$\,s &  $18$ & $144$ & $0.01$\,s  &  $134{,}352$ & $5{,}126$ & $8{,}017$   & $0.1$\,s \\
$9$  &  $12{,}501 + 9\phantom{0}$ &     $801$\,s &  $20$ & $180$  & $0.26$\,s  & $247{,}320$ & $7{,}198$ & $29{,}147$  & $0.6$\,s \\
$10$ & $19{,}300 + 10$ & $>3600$\,s &  $22$ & $220$  & $12.5$\,s  & $427{,}240$ & $9{,}773$ & $403{,}613$ & $11$\,s \\
\bottomrule
\end{tabular}
\end{table}

\subsection{Mycielski Graphs}

Mycielski's construction takes a graph $G$ and returns a graph $M(G)$
with the same clique number and chromatic number one
higher~\cite{mycielski:cm:1955}. Starting from $M_2 = K_2$, the
iterates $M_k = M(M_{k-1})$ are triangle-free with $\chi(M_k)=k$.
Refuting the $(k{-}1)$-coloring of $M_k$ is a classic hard family.
The textbook proof is an induction: $M_k$ consists of $M_{k-1}$, a
shadow vertex $u_v$ per vertex $v$ with $u_v$ adjacent to the neighbors of $v$, 
and an additional vertex $w$
adjacent to all shadows. If $w$ has color $a$, every shadow avoids
$a$, so recoloring each $a$-colored $v$ to its shadow's color turns a
$(k{-}1)$-coloring of $M_k$ into a $(k{-}2)$-coloring of $M_{k-1}$,
which does not exist by the same argument one level down.

This induction is a tower of lemmas, and it exercises every
distinctive feature of DRATL at once. Lemma $L_j$ has the coloring
constraints of $M_j$ as its antecedent clauses and a single \emph{empty}
consequent: it states that the $(j{-}1)$-coloring of $M_j$ is
contradictory, and an application under qualifier $Q$ concludes
$Q\lor\bot = Q$: a refutation made reusable by the qualifier
alone. The proof of $L_j$ defines the $j{-}1$ recolorings and derives
their coloring constraints (all RUP), then applies $L_{j-1}$ to each,
under the qualifier $\overline{z}_{w,a}$; each application concludes
$(\overline{z}_{w,a})$, and the empty clause follows from the
at-least-one clause for $w$. The lemmas require no bookkeeping
parameters: an application inside a lemma's proof is automatically
qualified by that proof's activation guard, so conclusions carry
exactly one caller guard at any nesting depth.

Nesting allows a single proof of each lemma to serve all of its
instantiations. Without sharing, each application of $L_j$ requires
a separate copy of its proof, including the $j{-}1$ applications of
$L_{j-1}$. Recursively expanding these applications yields a proof
whose size grows like $(k{-}1)!$, while the nested proof grows
linearly in the formula size up to a factor $k$
(Table~\ref{tab:mycielski}). Elaboration adds only about five percent
for the largest instance, preserving the sharing expressed in the
DRATL proof.

\begin{table}[t]
\caption{Mycielski graphs: the nested lemma tower. The first three columns show the graph, 
number of clauses and number of lemmas, respectively. ``DRATL'' shows the lines of the proof as written,
``elaborated'' and ``flat'' the additions of the expanded proof and
of the expansion without lemma sharing; ``check'' is the total
checking time: \texttt{dsr-trim} on the elaborated proof, then the
formally verified \texttt{srcheck} on its annotated
output.}\label{tab:mycielski}
\centering
\setlength{\tabcolsep}{8pt}
\begin{tabular}{crcrrrr}
\toprule
graph & clauses & lemmas & DRATL & elaborated & flat & check \\
\midrule
$M_4$ & $71$       & $2$ & $385$       & $463$       & $463$           & $0.03$\,s \\
$M_5$ & $307$      & $3$ & $2{,}324$   & $2{,}680$   & $3{,}761$       & $0.05$\,s \\
$M_6$ & $1{,}227$  & $4$ & $11{,}925$  & $13{,}221$  & $28{,}171$      & $0.26$\,s \\
$M_7$ & $4{,}625$  & $5$ & $55{,}048$  & $59{,}190$  & $211{,}033$     & $2.8$\,s \\
$M_8$ & $16{,}711$ & $6$ & $236{,}189$ & $248{,}343$ & $1{,}653{,}905$ & $32$\,s \\
\bottomrule
\end{tabular}
\end{table}

\section{Conclusions}

This paper demonstrates that hierarchy can be introduced into a
clausal proof framework to yield more compact proofs than the
flat proofs generated by existing solvers and preprocessors.
The ability to efficiently elaborate hierarchical proofs into flat
ones avoids the need for new proof frameworks or checking tools.  Our
results demonstrate the utility of this approach for several highly
structured challenge problems.

A clear next step in this effort is to have the elaborator also
generate the full proof details, yielding a proof in
the LRAT extension of the DRAT format~\cite{cruz:cade:2017}.  Such
a capability should be straightforward --- the lemma proofs are fully
self-contained, and deriving the hints for the steps outlined in
Figure~\ref{fig:compact} is mostly a matter of bookkeeping.  This
would enable a direct interface to multiple formally verified proof
checkers~\cite{codel:fmcad:2024,heule:itp:2017,tan:sttt:2023}.

The method of elaborating lemmas described in Figure~\ref{fig:compact}
could also be incorporated into other proof systems that allow the
introduction of extension variables.  These include clausal proof
systems that generalize
DRAT~\cite{codel:fmcad:2024,heule:nfm:2019,heule:jar:2019}, as well as
the VeriPB proof framework for pseudo-Boolean
reasoning~\cite{gocht:aaai:2021}.  As with the RAT rule, these other
proof systems include non-monotonic proof rules.  Some limitations may
need to be imposed on how these rules are used within the proof of a lemma
to ensure that no conflict arises with an instantiation.

Another direction for research is to incorporate lemma extraction into
a CDCL SAT solver, either as part of a preprocessing or an
inprocessing step.  The solver could identify multiple
instances of a common clausal pattern and introduce a lemma creating a
more useful representation of that pattern, similar to our processing
of clique constraints in the Queen graph coloring problem.

%
%
%

\newpage

%

\end{document}